\def \SAIT #1 #2 {{\em Mem.\ Soc.\ Astron.\ It.\/} {\bf #1}, #2}
\def \MESS #1 #2 {{\em The Messenger\/} {\bf #1}, #2}
\def \ASTRNACH #1 #2 {{\em Astron. Nach.\/} {\bf #1}, #2}
\def \AAP #1 #2 {{\em Astron. Astrophys.\/} {\bf #1}, #2}
\def \AAL #1 #2 {{\em Astron. Astrophys. Lett.\/} {\bf #1}, L#2}
\def \AAR #1 #2 {{\em Astron. Astrophys. Rev.\/} {\bf #1}, #2}
\def \AAS #1 #2 {{\em Astron. Astrophys. Suppl. Ser.\/} {\bf #1},
#2}
\def \AJ #1 #2 {{\em Astron. J.\/} {\bf #1}, #2}
\def \ANNREV #1 #2 {{\em Ann. Rev. Astron. Astrophys.\/} {\bf
#1}, #2}
\def \APJ #1 #2 {{\em Astrophys. J.\/} {\bf #1}, #2}
\def \APJL #1 #2 {{\em Astrophys. J. Lett.\/} {\bf #1}, L#2}
\def \APJS #1 #2 {{\em Astrophys. J. Suppl.\/} {\bf #1}, #2}
\def \APSS #1 #2 {{\em Astrophys. Space Sci.\/} {\bf #1}, #2}
\def \ASR #1 #2 {{\em Adv. Space Res.\/} {\bf #1}, #2}
\def \BAIC #1 #2 {{\em Bull. Astron. Inst. Czechosl.\/} {\bf #1},
#2}
\def \JSQRT #1 #2 {{\em J. Quant. Spectrosc. Radiat. Transfer\/}
{\bf #1}, #2}
\def \MN #1 #2 {{\em Mon. Not. R. Astr. Soc.\/} {\bf #1}, #2}
\def \MEM #1 #2 {{\em Mem. R. Astr. Soc.\/} {\bf #1}, #2}
\def \PLR #1 #2 {{\em Phys. Lett. Rev.\/} {\bf #1}, #2}
\def \PASJ #1 #2 {{\em Publ. Astron. Soc. Japan\/} {\bf #1}, #2}
\def \PASP #1 #2 {{\em Publ. Astr. Soc. Pacific\/} {\bf #1}, #2}
\def \NAT #1 #2 {{\em Nature\/} {\bf #1}, #2}

\documentstyle{memsait}
\input epsf.sty
\begin{opening}
\title{Why is the X-ray luminosity function different for the
 Hyades and Praesepe?}
\author{D. Barrado y Navascu\'es$^1$, J.R. Stauffer$^1$, S.
Randich$^{2,3}$}
\institute{$^1$Center for Astrophysics, Cambridge, USA\\
$^2$ESO, Garching, Germany\\
$^3$Osservatorio di Arcetri, Florence, Italy}
\date{} 
\end{opening}

\begin{document}

\oddpagefooter{}{}{} 
\evenpagefooter{}{}{} 
\ 
\bigskip

\begin{abstract}
We have studied dF-dK and dM stars belonging to Praesepe in order
to determine
 membership via radial velocities  and to compare  several
stellar
properties, such as X--ray luminosities, H$\alpha$ equivalent
width, etc, 
between this cluster and Hyades, which have similar age. We show
that all properties are analogous in both clusters except the
distribution 
of Lx. This fact could be related to difficulties in the analysis
of the Lx 
upper limits or to more fundamental 
reasons, such as binarity rates and differences in the properties
of the binaries of each cluster.
\end{abstract}

\section{Introduction}

Randich \& Schmitt (1995), using ROSAT PSPC data,
 have studied the X-ray properties of the Praesepe
cluster and have shown that the X-ray luminosity function
is very different from  that corresponding to the Hyades, which
has a similar age and  metallicity. This result seems to 
contradict our assumptions about the evolution of stellar
rotation and the age-rotation-activity paradigm.
However, Mermilliod (1997a) has established that the rotational
velocity distribution is very similar for both clusters.
The goal of this paper is to try to disentangle the problem and
explain the apparent dichotomy of these coeval clusters.

The observations analyzed in this paper were collected during Jan
11--13, 1995,  and Jan 27--28, 1996, at the MMT. The first
observing run was devoted  to late dF-dK Praesepe stars, in order 
to measure their radial velocities (with a resolution of 0.2
\AA). During the second run, we obtained radial velocities and
H$\alpha$ equivalent widths of Praesepe dM members
(resolution$\sim$1.5 \AA). Figure 1a,b shows both 
color-magnitude diagrams for these samples (solid symbols for the
targets).

\section{dF-dK stars in Praesepe}

In order to determine whether non--members significantly
contaminate
the Praesepe catalog used by Randich \& Schmitt (1995), we have
measured  radial velocities of a subsample which contains 
16 stars having  X--ray detections and 21 stars with X-ray upper
limits. 22 of these stars have radial velocities consistent 
with that expected for single Praesepe members. Of the other 15
stars, 8 have been shown to be spectroscopic binary members by 
extensive  radial velocity studies (Mermilliod 1997b). 
One star, KW460, is a probable non-member.
Based on incomplete information about
photometry, lithium abundances, proper motions, etc, we believe
that the 
remaining 6 stars are also binaries and members of the cluster.
Our conclusion is that the Praesepe sample is not significantly
contaminated by non-members. Therefore, we must search elsewhere 
to explain the apparent difference between the X-ray detection 
frequency in binaries of Praesepe (56\%) and Hyades (96\%,
 Stern et al. 1995).

Figure 2a,b shows Lx against (B--V) color for dG
Praesepe and Hyades stars (detection are shown as solid circles,
whereas upper limits appear as open triangles, the solid lines
represent where the lowest X-ray detection lies in each cluster).
Proper motions, photometry and radial velocities prove that there
are not a significant number of  spurious members in the Praesepe
dF--dK sample.
Therefore, the obvious differences in the X-ray distribution seem
real. We emphasize  the difference in the  distributions
  below Log~Lx=28.4. Praesepe
has 12 stars with luminosities under this value, whereas Hyades
only has 1. We have analyzed a deep pointing, retrieved from the
the ROSAT archive, to verify if
the previous upper limits  in Praesepe were assigned correctly.
We detected the same stars which were
detected in the raster scan (and did not detect those that were not
detected). The ULs which we got for undetected stars are consistent
with those from the raster, although we cannot exclude that some of the
latter were underestimated by a factor of ~2. This, however, would not
be enough to fully explain the discrepancy between Hyades and Praesepe
dG XLDF. This situation leads to the interpretation
that the  different X-ray properties of dF--dK members of both
clusters could be related with the binarity rates and different
distributions of the orbital elements, such as the orbital
period.
Fig. 2a,b suggests that the sensitivity of the Hyades  and the
Praesepe data could be different, and this fact would also  help to
solve, partially, the problem.


\begin{figure}
\vspace{4.5cm}
\includegraphics{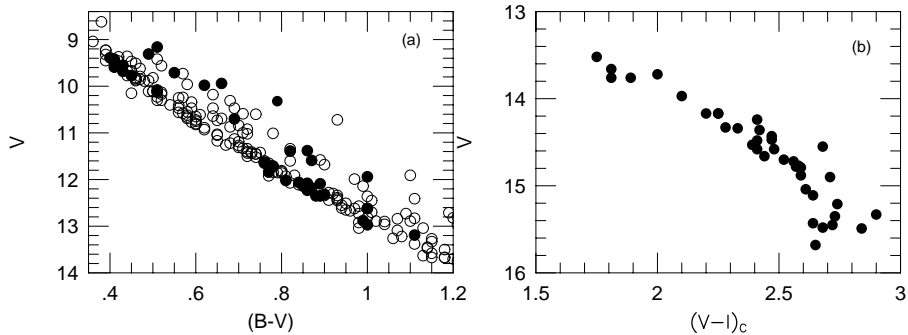}
\caption[h]{Color magnitude diagrams for the dF-dK (a) and
dM (b) Praesepe stars. Our targets are shown as solid circles.}
\end{figure}


\section{dM stars}

The difference between the X--ray luminosity function of dF--dK 
members of the two clusters extends also to the M dwarfs,
 as shown in  Figure 2c,d. X-ray detections and 
upper limits are represented as circles and triangles,
respectively; the  strong and weak H$\alpha$ features, as defined
from Figure 3 are shown as solid and open symbols, respectively.
Note that none of our dM Praesepe stars were detected in X--rays.
There is a clear correspondence between strong H$\alpha$ emission
and detection in X-rays in the case of Hyades, as expected.
However, this relation does not appear to hold for  Praesepe.
Although it is possible to understand the fact that there are
stars with high X-ray upper limit and weak H$\alpha$ (after all,
the actual Lx can  be much smaller than the upper limit),
it is difficult to explain the presence of several stars having
strong H$\alpha$ emission but, simultaneously, very low X--ray
upper limits. Figure 3a,b, where the H$\alpha$ equivalent width
(our data and values from William et al. 1994 are included)  is
represented against the (V-I)$_{\rm C}$ color, shows that there
is no obvious difference in the H$\alpha$ distribution, except
that several Praesepe M dwarfs are quite active. However, the
situation in X-rays is the opposite. Again, the factor 2 in the
re-assignation of the Lx upper limits could help, but it would not
be enough to explain all the differences, in particular the lack
of correspondence  between H$\alpha$ and Lx (stars having strong
H$\alpha$ and low Lx upper limits). 


\begin{figure}
\vspace{7.8cm}
\includegraphics{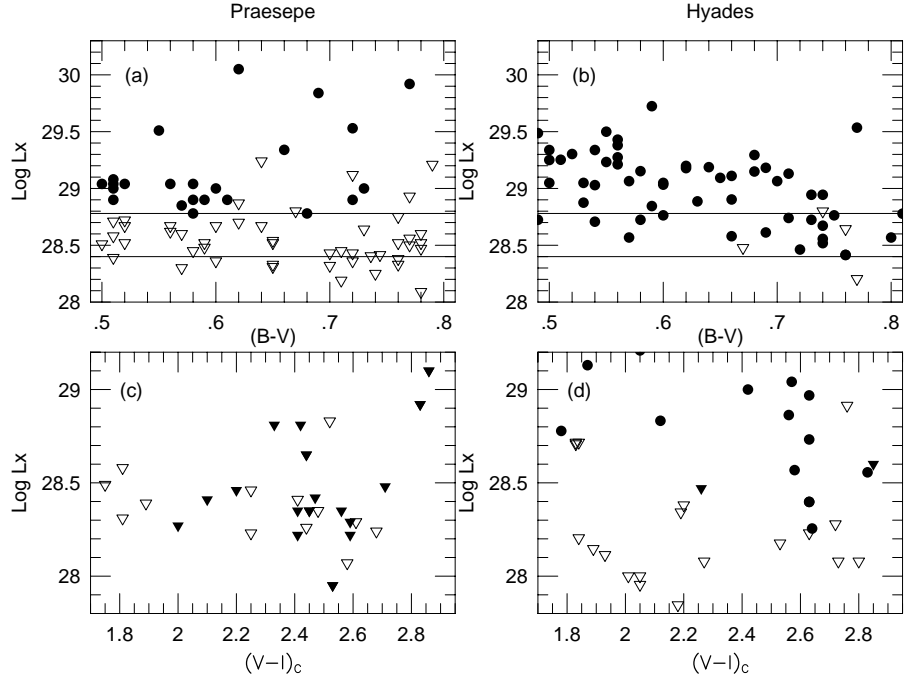}
\caption[h]{X-ray luminosities against color indices. a,c
Praesepe;
b,d Hyades. See text.}
\end{figure}

\section{Conclusions}

We have tried to understand the reasons for the different
X-ray properties of late type stars in the Hyades and Praesepe. 
We have studied two different samples of stars: dF-dK Praesepe
members for which we have derived radial velocities
 and dM Praesepe stars for which we have determined chromospheric
activity levels. We have shown that the Praesepe members catalog
has few or no spurious members and so the inclusion of
non--members
is not the cause for Praesepe's apparent lower X--ray activity.
 Other stellar
activity indicators, such as H$\alpha$, have the same
distribution in both clusters, as do other properties such as
the rotational velocities and the lithium abundances. That is,
the only difference is the X--ray distribution.

One possibility is that the X--ray upper limits of Praesepe stars
have been underestimated. Our new analysis shows that they could
be increased only a factor 2, and we think it is
unlikely that this  can be the complete explanation.
Alternatively, there may be real intrinsic
difference between some additional property --the binary
frequency or
metallicity, for example-- which could contribute to the 
dichotomy in the coronal properties. Further optical spectroscopy
of Praesepe may help to answer this question. A definitive answer
could 
be obtained by new X--ray observations once AXAF or XMM are launched.

\begin{figure}
\vspace{4.5cm}
\includegraphics{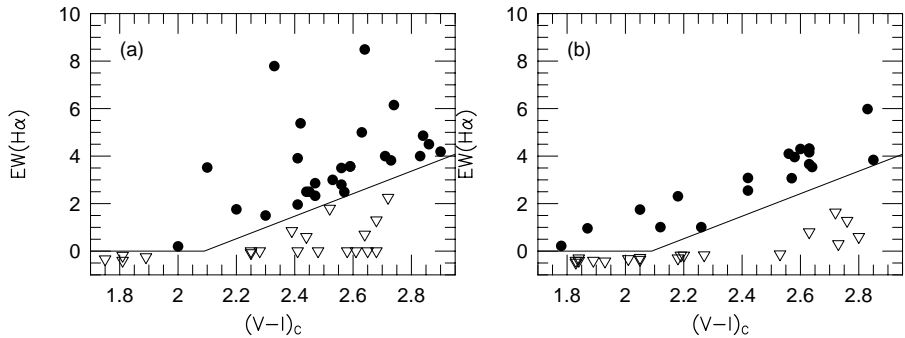}
\caption[h]{EW(H$\alpha$) against (V--I)$_{\rm C}$
color for Praesepe (a) and Hyades (b). The solid line
separates strong and weak H$\alpha$, as discussed in the text.} 
\end{figure}


\acknowledgements
DBN thanks the Real Colegio
Complutense at Harvard University and the MEC/Fulbright program
for their fellowships.
 JRS acknowledges support from NASA Grant
NAGW-2698.


\begin{thebibliography}{ }














\item Mermilliod, J-C., 1997a, this issue 

\item Mermilliod, J-C., 1997b, priv. comm.

\item Randich, S., Schmitt, J. H. M.
M.,  1995, \AAP 298, 115

\item Stern, R. A., Schmitt, J. M.
H. H., Kahabka, P. T.,  1995, \APJ 448, 683

\item Williams, S. D., Stauffer, J. R., Prosser,
C. F., Herter, T., 1994, \PASP 106, 817



\end{thebibliography}
\end{document}